\documentstyle[epsfig,twocolumn,aps,amssymb]{revtex}
\begin{document}
\draft

\title{Damping and revivals of collective oscillations in a finite-temperature
 model of trapped Bose-Einstein condensation}
\author{B. Jackson and C. S. Adams}
\address{Department of Physics, Rochester Building, University of Durham, 
South Road, Durham, DH1 3LE, UK.}
\date{\today}
\maketitle
\begin{abstract}
 We utilize a two-gas model to simulate collective oscillations of a 
 Bose-Einstein condensate at finite temperatures. The condensate is described
 using a generalized Gross-Pitaevskii equation, which is coupled to a thermal
 cloud modelled by a Monte Carlo algorithm. This allows us to
 include the collective dynamics of both the condensed and non-condensed
 components self-consistently. We simulate quadrupolar excitations, and 
 measure the damping rate and frequency as a function of temperature. We
 also observe revivals in condensate oscillations at high temperatures, and
 in the thermal cloud at low temperature. Extensions of the model to include
 non-equilibrium effects and describe more complex phenomena are discussed.
\end{abstract}

The first experimental observation of Bose-Einstein condensation (BEC) in
magnetically trapped alkali atoms in 1995 \cite{anderson95,davis95,bradley95}
was a precursor to an explosion of interest in the properties of 
weakly-interacting Bose gases. Much of the subsequent theory \cite{dalfovo99}
has focused on the dynamics of the condensate, including phenomena such as 
collective excitations and vortex motion. In the limit of zero temperature, 
one can represent the 
condensate by a macroscopic wavefunction analogous to a classical 
field. In this case the behavior can be described in terms of the 
Gross-Pitaevskii (GP) equation, which has the form of a nonlinear 
Shr\"{o}dinger
equation. Extension of the description to finite temperatures, where one must
include fluctuations upon the condensate wavefunction, is a considerable 
challenge. However, the motivation is clear, as such a description would
allow direct comparison with experiments where a non-condensed thermal cloud
is present, as well as revealing new phenomena such as damping 
of collective modes 
\cite{pitaevskii97,liu97,fedichev98,giorgini98,jin97,stamperkurn98} and the
decay of metastable vortices \cite{rokhsar97,fedichev99}.

Amongst the most compelling evidence for the validity of the GP equation at
low temperatures is its quantitative agreement with experiment for low-energy
collective modes. However, consistent theoretical descriptions at higher
temperatures have proved far more elusive, where experiments have demonstrated
marked frequency shifts and damping of the condensate modes in the presence
of a significant non-condensed component \cite{jin97,stamperkurn98}.
Theoretical studies have tended to concentrate on one of two regimes,
depending upon the density and temperature of the system. At high densities,
where collisions are sufficiently rapid to force the system into local
equilibrium, the dynamics of the condensate and thermal cloud can be described
by a set of coupled hydrodynamical equations 
\cite{zaremba98,nikuni99,zaremba99}.
Damping mechanisms in this case are of a dissipative type (i.e.\ viscosity and
thermal relaxation). For very dilute systems or at low temperatures the
mean free path of the elementary excitations become comparable to the size
of the system and collisions play only a minor role. Damping in this
collisionless
regime is not related to thermalization processes but to coupling between
excitations, and can be described within the framework of mean-field
theories \cite{giorgini98}. The collisionless regime may be
appropriate for the JILA experiments \cite{jin97}, while the MIT experiments
lie between the collisionless and hydrodynamical regimes \cite{stamperkurn98}.

Here we present a simple model of a finite-temperature BEC system, and use
numerical simulations to find the temperature-dependent frequency and
damping of a quadrupole mode. Essentially, we utilize a two-fluid approach,
where the ground-state condensate and low-energy, highly-occupied `classical'
modes are described by a generalized GP equation, while the thermal cloud,
which is composed of higher-energy excitations, is simulated using a
Monte Carlo approach. One advantage of this model is that we do not need
to invoke strong assumptions about particle collision times, so that we can
study the intermediate region between the collisionless and hydrodynamical
regimes. We also consistently include the collective dynamics of the thermal
cloud, which are particularly significant at temperatures near to the Bose
transition. This aspect is often absent from other treatments, for example
frequency calculations from solving Hartree-Fock-Bogoliubov (HFB) equations
\cite{hutchinson97,dodd98}. Finally, the model can potentially be extended to
simulate more complex situations, such as vortex decay and response to
time-dependent probes.

The generalized GP equation for the condensate wavefunction,
$\Psi ({\mathbf{r}},t)$, is written in the Popov approximation (i.e.\ the
`anomalous' density, $\tilde{m}=0$) as \cite{zaremba99}
\begin{eqnarray}
 i\hbar \frac{\partial}{\partial t} \Psi ({\mathbf{r}},t) = \Bigg( 
 -\frac{\hbar^2}{2m} \nabla^2 + V ({\mathbf{r}},t) + g
 [2\tilde{n} ({\mathbf{r}},t) + \nonumber \\
 n_c ({\mathbf{r}},t)]-i\Lambda ({\mathbf{r}},t) \Bigg) \Psi ({\mathbf{r}},t),
\label{eq:GP-gendamp}
\end{eqnarray}
where $\tilde{n} ({\mathbf{r}},t)$ is the non-condensate density, while
$n_c({\mathbf{r}},t) =|\Psi({\mathbf{r}},t)|^2$ is the condensate density
(where the wavefunction is normalized to the number of condensate atoms,
$N_c$). The condensate is subject to an external trap potential,
$V({\mathbf{r}},t)=m(\omega_r^2 r^2+\omega_z^2 z^2)/2$, as well as mean-field
effects arising from the
thermal cloud and the remainder of the condensate. These interactions are
parameterized by the coupling constant, $g=4\pi \hbar^2 a/m$, where $m$ is
the atomic mass and $a$ is the s-wave scattering length. In this paper
we shall concentrate on $^{87} {\rm Rb}$, where $a=5.5\,{\rm nm}$.
The dissipative term, $\Lambda({\mathbf{r}},t)$, represents
collisions that transfer atoms between the condensed and non-condensed
components. Here we shall assume local equilibrium between the two components,
so that this term vanishes, $\Lambda=0$. Generalization of this model to 
include this effect will be the subject of future work.

Given the non-condensate density $\tilde{n}$ as a function of position and
time,
the GP equation (\ref{eq:GP-gendamp}) can readily be solved using techniques
discussed in our previous work \cite{jackson00}. Briefly, the GP equation is 
re-scaled for convenience, before being propagated over a small time-step
$\Delta t$ using a FFT method. The time-independent problem, corresponding
to finding initial conditions of our simulations, can be solved by propagating
in imaginary time, so that an arbitrary wavefunction quickly diffuses to
the ground state solution. The equilibrium thermal distribution can be
calculated under a semi-classical approximation \cite{giorgini97}, where in
a harmonic trap of mean frequency $\bar{\omega}$ the discrete energy levels
are replaced by a continuous function
$\epsilon_{\rm HF}=p^2/2m + V({\mathbf{r}})+2g[n_c({\mathbf{r}})+\tilde{n}
({\mathbf{r}})]-\mu$. This corresponds to the energy of a single particle
moving within the mean-field. The semi-classical approximation is valid
under the condition that $k_B T \gg \hbar \bar{\omega}$, and when the
number of trapped atoms is large \cite{dalfovo99}. An integration over
momentum states then simply yields
\begin{equation}
 \tilde{n} ({\mathbf{r}}) = \frac{1}{\lambda_T^3} g_{3/2} (z),
\label{eq:ndens}
\end{equation}
where $z=\exp [-\beta \{ V({\mathbf{r}})+2g(n_c({\mathbf{r}})+\tilde{n}
({\mathbf{r}}))-\mu \} ]$ is the fugacity,
$\lambda_T = (2\pi\hbar^2/mk_B T)^{1/2}$ is the thermal wavelength, and
$g_{\alpha}(z) = \sum^{\infty}_{l=1} z^l / l^{\alpha}$. The total number of
atoms in the system is then given by $N=N_c+\int {\rm d}{\mathbf{r}}\,
\tilde{n}=N_c+\tilde{N}$, where $\tilde{N}$ is the number of atoms
in the thermal cloud.

Self-consistent solution of (\ref{eq:GP-gendamp}) and (\ref{eq:ndens}) yields
good approximations for the condensate and non-condensate
densities at equilibrium. In particular, we iterate using successive 
evaluations of (\ref{eq:ndens}) and imaginary time propagation of the 
condensate wavefunction, to find the densities as a function of $N$ and $T$.
Given this initial condition, the system dynamics resulting from a varying 
trap potential can be found 
by solving the time-dependent GP equation using propagation in real time.
However, the problem of finding the non-condensate density becomes more
challenging as this is also time-dependent. Under the semi-classical
and Hartree-Fock approximations, and the assumption that the effective 
potential 
$U_{\rm eff} ({\mathbf{r}},t)=V({\mathbf{r}},t)+2g[n_c({\mathbf{r}},t)+
\tilde{n} ({\mathbf{r}},t)]$ varies slowly in space,
one can show \cite{zaremba99,kadanoff89}
that the cloud of quasiparticle excitations may be described using a
Boltzmann kinetic equation
\begin{eqnarray}
 \frac{\partial f({\mathbf{p}},{\mathbf{r}},t)}{\partial t} + 
 \frac{{\mathbf{p}}}{m} \cdot \nabla f({\mathbf{p}},{\mathbf{r}},t) 
 - \nabla U_{\rm eff} \cdot \nabla_{\mathbf{p}} 
 f({\mathbf{p}},{\mathbf{r}},t) \nonumber \\
 = \left. \frac{\partial f}{\partial t} \right|_{\rm coll}.
\label{eq:kinetic}
\end{eqnarray}
The relationship between the phase-space distribution function
$f({\mathbf{r}},{\mathbf{p}},t)$ and the non-condensate density is simply
given by
\begin{equation}
 \tilde{n} ({\mathbf{r}},t) = \int \frac{{\rm d}{\mathbf{p}}}{(2\pi\hbar)^3}
 f ({\mathbf{p}},{\mathbf{r}},t).
\label{eq:noncond}
\end{equation}
The right-hand term of (\ref{eq:kinetic}), which provides the scattering rate 
of state ${\mathbf{p}}$,
is given by an integral representing two-body collisions between atoms in the
thermal cloud within the Born approximation
\begin{eqnarray}
 \left. \frac{\partial f}{\partial t} \right|_{\rm coll} =
 \frac{2g^2}{(2\pi)^5 \hbar^7} \int {\rm d}{\mathbf{p}}_2 \int
 {\rm d} {\mathbf{p}}_3 \int {\rm d}{\mathbf{p}}_4 \nonumber \\
 \delta({\mathbf{p}} +{\mathbf{p}}_2 - {\mathbf{p}}_3 - {\mathbf{p}}_4)
 \delta (\tilde{\epsilon}_p + \tilde{\epsilon}_{p_2} - 
 \tilde{\epsilon}_{p_3} - \tilde{\epsilon}_{p_4}) \nonumber \\
 \times[(1+f)(1+f_2)f_3 f_4 - f f_2 (1+f_3) (1+f_4)],
\label{eq:col22}
\end{eqnarray}
with $f \equiv f({\mathbf{p}},{\mathbf{r}},t)$ and 
$f_i \equiv f({\mathbf{p_i}},{\mathbf{r}},t)$. Locally, an excited atom has
the HF energy $\tilde{\epsilon}_p ({\mathbf{r}},t) = p^2/2m +U_{\rm eff} 
({\mathbf{r}},t)$. 
As above, we neglect collisions that transfer atoms
between the two components. The collision integral (\ref{eq:col22}) differs
from that of a classical gas \cite{huang87} by the inclusion of $(1+f)$
factors that represent Bose enhancement of scattering into occupied states.

In Refs.\ \cite{zaremba98,nikuni99,zaremba99}, moments of the
kinetic equation (\ref{eq:kinetic}) were taken to yield hydrodynamical 
equations
\cite{huang87}. These can be solved explicitly under certain conditions using
a variational method \cite{zaremba99}. An alternative approach is to solve
(\ref{eq:kinetic}) directly. In general, this is very difficult owing to the
six-dimensional nature of phase space. One possibility is to work under 
the {\it assumption of sufficient ergodicity}. Ergodicity assumes
that the distribution of atoms in phase space depends only on their energy,
$\tilde{\epsilon}$.
Then (\ref{eq:kinetic}) reduces to an equation of motion for 
$f(\tilde{\epsilon})$.
This assumption is well-known in the literature and has been used to model
evaporative cooling \cite{luiten96,holland97} and condensate growth 
\cite{snoke89,semikoz95,bijlsma00} in Bose gases. However, ergodicity assumes
that any deformation in momentum or position space is isotropic, or that the
ergodic mixing time is shorter than the elastic collision time. In general
non-equilibrium situations this assumption is not valid. In addition, we are
primarily interested in the gas dynamics in position space and its coupling 
to the condensate. Hence, a Monte Carlo technique \cite{lopez98,wu96,wu97}
is more appropriate here. In particular, we utilize a direct simulation
Monte Carlo (DSMC) method, as performed to model evaporative cooling 
in Bose gases \cite{wu96,wu97}, and described in detail for classical 
gas dynamics in \cite{bird94}. We now discuss our extension of this
model to simulate the thermal cloud coupled to the condensate.  

The direct simulation Monte Carlo (DSMC) method was first developed by Bird to 
describe classical gas flows \cite{bird94}. It is equivalent to solving the 
Boltzmann equation in phase-space, except that it recognizes the discrete
nature of the gas on a microscopic level. In principle, the trajectory of
each atom could be followed at all times, so that the state of the system
is completely described by storing $({\mathbf{r}},{\mathbf{v}})$ for all
atoms. However, the calculation becomes unfeasible in the presence of 
interparticle collisions. Bird's method makes the key assumption that the
free particle motion and collisions are uncoupled over a short time interval,
$\Delta t$. This provides an accurate description of the gas so long that
$\Delta t \ll \tau_{\rm coll}$, where $\tau_{\rm coll}$ is the mean collision
time. Hence the DSMC method is most appropriate for describing gases in the
Knudsen regime, where the mean free path is much larger
than the size of the system. The technique is therefore well suited to dilute
alkali gases.

First, the atoms are moved over distances appropriate to their velocity
components, ${\mathbf{v}}_k$ ($k \in \{ x,y,z \}$), such that 
${\mathbf{r}}_{k+1}={\mathbf{r}}_k+{\mathbf{v}}_k \Delta t$, before collisions
are treated. To ensure that collisions only take
place between near neighbours, position space is divided into cubic cells of a
size much smaller than the dimensions of the cloud. The 
number of atoms is counted
in each cell to furnish the local density $\tilde{n} ({\mathbf{r}})$. Pairs
of atoms in a cell are then chosen at random, and the momenta after a collision
is calculated {\it a priori}. To account for energy and momentum conservation 
the collision is most conveniently treated in the atomic centre-of-mass frame.
Two further random numbers, $R_1$ and $R_2$, are chosen to determine
the scattering angles $\phi=2\pi R_1$ and $\cos \theta =1-2R_2$, where
$R_{1,2} \in [0,1]$. To decide whether the atoms actually do collide, the
following algorithm is used. First, the mean number of collisions
locally in time $\Delta t$ is calculated using
\begin{equation}
 \bar{\rho} ({\mathbf{r}},t)=\tilde{n} ({\mathbf{r}},t) \sigma v_r \Delta t
 [1+f({\mathbf{p}}_3,{\mathbf{r}},t)] [1+f({\mathbf{p}}_4,{\mathbf{r}},t)],
\label{eq:coll-no}     
\end{equation}
where $\sigma=8\pi a^2$ is the scattering cross-section for bosons in a hard
sphere model, and $v_r=|{\mathbf{v}}_2-{\mathbf{v}}_1|$ is the relative 
velocity of the two atoms. A `quantum scattering factor' is also included which
represents the effect of Bose statistics, where ${\mathbf{p}}_3$ and 
${\mathbf{p}}_4$ are the momenta of the collision products. To estimate the
distribution function, the number of atoms are counted within subcells in
momentum space, which in turn are subdivisions of the positional cells.
Strictly speaking, each phase-space subcell should have a volume of 
$h^3$, which is the minimum value allowed by the uncertainty principle. 
However, the computational time required to sort the atoms increases linearly 
with the total number of subcells, and can become prohibitively large without 
some form of coarse graining. We therefore count the number of atoms 
${\mathcal{N}}_{\rm sc}$ within larger subcells, which is renormalized to
yield $f({\mathbf{p}},{\mathbf{r}},t) \simeq {\mathcal{N}}_{\rm sc} h^3 /
{\mathcal{V}}_{\rm p} {\mathcal{V}}_{\rm r}$, 
where ${\mathcal{V}}_{\rm p}$ and ${\mathcal{V}}_{\rm r}$ are the volumes of
cells in momentum and position space respectively. We find that our results
are largely independent of the number of cells and subcells for sufficiently 
large numbers. For example, for the computations described below we use
8000 position cells subdivided into 9261 momentum subcells.

As $\bar{\rho} ({\mathbf{r}},t) \ll 1$, the collision probability is
given by $P_{\rm coll} = 1-{\rm e}^{-\bar{\rho} ({\mathbf{r}},t)}$.
A further random number, $R_0 \in [0,1]$, is compared to $P_{\rm coll}$. Only 
if $R_0 < P_{\rm coll}$ does a collision takes place. 
Once this procedure has been repeated for all of the atoms in each cell, the
final part of the time-step involves updating the atom velocities to 
account for gradients in the external potential, ${\mathbf{v}}_{k+1}=
{\mathbf{v}}_k + \Delta {\mathbf{v}}_k$, where
$\Delta {\mathbf{v}}_k = - \partial_k U_{\rm eff} \Delta t/m$.

To simulate the coupled system, alternating Monte Carlo and GP propagation 
steps are performed during each time-step, $\Delta t$, where thermal atoms
move in the mean-field potential, $U_{\rm eff}$. The thermal gas density is 
calculated at all points by counting atoms in each
cell. The cells do not necessarily correspond to the GP grid, so cubic spline
interpolation is used to smooth $\tilde{n}$. This prevents discontinuities 
in the mean-field potential, that may lead to instabilities in the FFT method 
used to propagate the condensate wavefunction. 

The first stage of the simulation is to find the initial state for a 
prescribed temperature, $T$. The numbers of condensate and thermal atoms are
found by the semi-classical algorithm described above. The 
equilibrium condensate density evaluated by this method, as well as a 
Maxwell-Boltzmann distribution for the thermal atoms, are used as an initial
state for the MC-GP algorithm. The condensate is propagated through imaginary 
time while the thermal cloud relaxes to equilibrium. A time-dependent
trap potential $V({\mathbf{r}},t)$ is applied and the system allowed
to propagate in real time.

\begin{figure}
\centering
\psfig{file=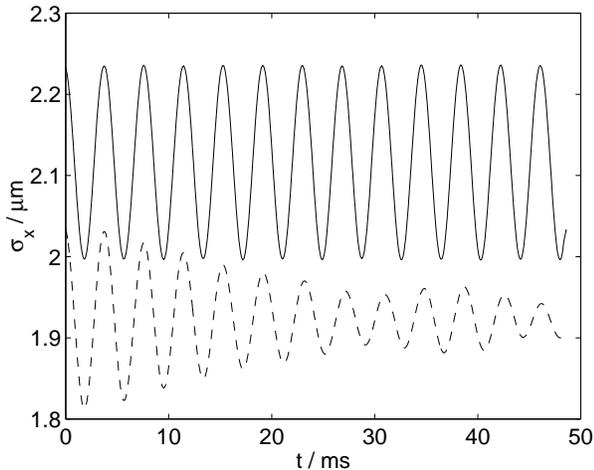, scale=0.6, angle=0}
\caption{\label{dampfig1} Quadrupole ($l=2$, $m=0$) oscillations of a
 condensate at $T=20\, {\rm nK}$ (solid) and $T=160\, {\rm nK}$ (dashed). The
 width of the condensate is represented by the standard deviation along $x$, 
 $\sigma_x$. Damping is observed at higher temperatures due to coupling with 
 the non-condensed thermal cloud.}
\end{figure}

We simulate collective excitations of $N=40000$ atoms in a disk-shaped trap
($\omega_r=2\pi\times 131\, {\rm Hz}$, $\omega_z=\sqrt{8}\omega_r$).
These are similar parameters to the JILA experiment \cite{jin97}. We 
study the $m=0$ quadrupole mode, which is excited by a sudden $10\%$ increase 
in the radial frequency, $\omega_r$ \cite{softer}. 
The subsequent condensate oscillations are shown in 
Fig.\ \ref{dampfig1}. We clearly observe damping at higher temperatures, which
is absent at $T=20\, {\rm nK}$. Note that we determine the widths of both
the components by calculating the standard deviation 
$\sigma_k=\sqrt{\langle k^2 \rangle - \langle k \rangle ^2}$, where we use
$\langle k^n \rangle =\int {\rm d}^3 {\mathbf{r}}\: k^n |\Psi|^2$ for the
condensate and $\langle k^n \rangle = \sum_i k_i^n / N$ for the thermal cloud.
To avoid large statistical fluctuations in the thermal component, especially 
at low temperatures, we simulate ten times the physical number of atoms 
(equivalent to repeatedly running our simulations: a time-consuming process). 
For consistency the density and phase-space density of the gas are rescaled 
appropriately. 

\begin{figure}
\centering
\psfig{figure=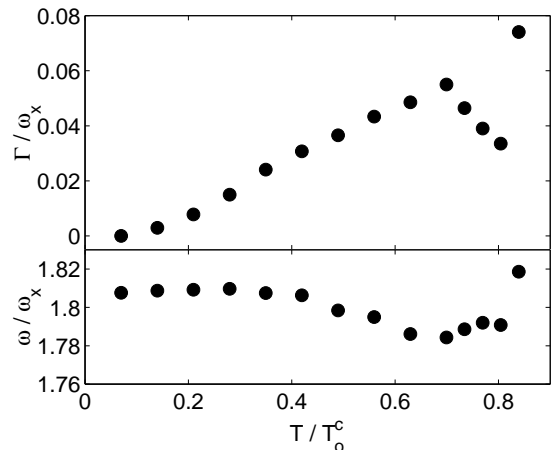, scale=0.6, angle=0}
\caption{\label{dampfig2} The damping rate, $\Gamma$ (top) and frequency 
 $\omega$ (bottom) of quadrupole $m=0$ oscillations in a cloud of 40000 atoms.
 The $x$-axis is 
 plotted as function of temperature, $T/T_c^0$, where $T_c^0=286\, {\rm nK}$ 
 is the ideal critical temperature, while the $y$-axes are plotted with 
 respect to the trap frequency $\omega_x=2\pi \times 131\, {\rm Hz}$.}
\end{figure}

We fit the condensate widths along $x$ and $y$ to an exponentially decaying 
sine function: $\sigma_k (t) = A {\rm e}^{-\Gamma t} \cos \omega t + B$.
The oscillation decay rate, $\Gamma$, and frequency, $\omega$,  are plotted 
against temperature in Fig.\ \ref{dampfig2}. The damping increases
from zero at low temperatures, before tending towards a linear dependence at
intermediate values ($T<0.7T_c^0$). This is in agreement with the expected
behaviour of Landau damping in homogeneous and trapped condensates, 
where in the limit of zero temperature
the damping has a $\Gamma \sim T^4$ dependence, while at higher temperatures
$\Gamma \sim T$ \cite{pitaevskii97,fedichev98,giorgini98,guilleumas00}. We 
also observe quantitative agreement with previous
theory \cite{fedichev98,guilleumas00,reidl00} and experiment in this regime. 
For example, at 
$T=200\, {\rm nK}$ ($T/T_c^0 \simeq 
0.7$) we find that $\Gamma \simeq 45.3\, {\rm s}^{-1}$, in fair agreement
with the experimental value of $90\pm40\, {\rm s}^{-1}$ \cite{jin97}. 
Landau damping arises due to mean-field 
coupling between fluctuations in the condensate wavefunction, $\delta \Psi$, 
and in the non-condensate density, $\delta \tilde{n}$ \cite{giorgini98}. 
Physically this is 
equivalent to absorption of a quantum of the collective mode by a thermal 
excitation \cite{liu97,fedichev98}. 
We find no damping at low temperatures, so that Balieav damping, which is 
active even at zero temperature, is not observed. This is to be expected, as 
the mechanism involves the decay of the collective mode into two lower 
frequency excitations and is equivalent to coupling
between $\delta \Psi$ and fluctuations in the anomalous density, 
$\delta \tilde{m}$ \cite{giorgini98}, which are neglected in our model. 
Nevertheless, the model is still consistent because Balieav damping is 
expected to be suppressed for the lowest modes in trapped condensates due to 
the discrete nature of the levels at low energy.  

Note that the Landau damping observed here should not be confused with the 
damping mechanism discussed in
\cite{nikuni99,zaremba99}, which is due to collisions that excite 
condensate atoms into the thermal cloud. For example, Landau
damping is a three excitation process as opposed to the four excitations 
involved in collisional damping. Our approach is justified as a first 
approximation because for relevant parameters the magnitude of collisional 
damping is significantly smaller than Landau damping \cite{williams00}.

\begin{figure}
\centering
\psfig{figure=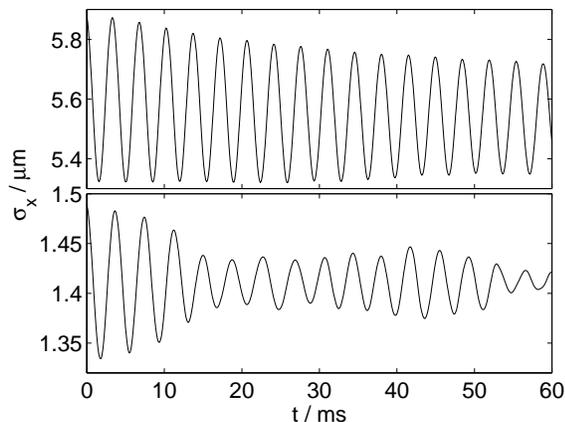,scale=0.6, angle=0}
\caption{\label{beat} Quadrupolar oscillations in the thermal cloud (top) and 
 condensate (bottom) at $T=240\, {\rm nK}$. A revival of the condensate 
 oscillations occurs at $t\simeq 40\, {\rm ms}$. }
\end{figure}

We observe a dip in the damping rate in the region $0.7T_c^0 < T < 0.8T_c^0$.
This is related to an interesting `beating' effect in the condensate 
oscillation. In this temperature range the oscillations are seen to damp
rapidly at early times, before reviving at a smaller amplitude after 
approximately ten oscillations. As a result, the fitting function tends to
underestimate the damping rate. As shown in Fig.\ \ref{beat}, the condensed
and normal components oscillate at slightly different frequencies due to 
their weak coupling, and the condensate is much more highly damped than 
the thermal gas. The latter is a consequence of the more massive thermal
cloud at this temperature (so that the back-action from Landau damping has 
less of an impact) and 
the small `internal' damping of the cloud from thermalization processes. As a 
result the thermal cloud acts as a kind of energy reservoir. When the 
oscillations of the two components are in anti-phase the condensate 
oscillations are strongly damped; however, when they are in phase the thermal
cloud tends to drive the condensate oscillations. This beating effect is
most noticeable when one component is significantly larger than the other.
Correspondingly, we see the same effect in the thermal cloud at low 
temperatures.    

The condensate oscillation frequencies are also plotted in Fig.\ 
\ref{dampfig2}.
The figure shows a small downward shift in frequency for $T<0.6T_c^0$
\cite{jin97,stamperkurn98,dodd98,reidl00}. An increase in frequency above 
$0.6T_c^0$ is also
observed; however, this is not as large as that seen in the JILA experiment
\cite{jin97}, where the frequency approaches $2\omega_x$ in this region. A 
possible explanation for the experimental behaviour was provided by Bijlsma and
Stoof \cite{bijlsma99}, who suggested a cross-over between normal modes where 
the two components oscillate in phase and anti-phase. However, as noted
previously we find that the components oscillate at slightly different
frequencies, and this description is inappropriate here. We may be able to 
see this effect at higher temperatures, though unfortunately the condensate
in this regime is small and more sensitive to local fluctuations in the
thermal cloud density, leading to unacceptable errors. The experimental
data also suffers from large errors in this region, making a direct
comparison difficult.    

To summarize, we have studied frequency shifts and Landau 
damping due to mean-field coupling between the condensate and the thermal 
cloud. We also observe revivals in the condensate oscillations at high 
temperature due to back-coupling from the thermal cloud, illustrating that
damping is not completely irreversible. Future extensions to our 
Monte Carlo simulations could include collisions 
that excite atoms out of the condensate \cite{nikuni99,zaremba99,williams00}. 
Along with a description of additional damping processes in collective 
excitations, this extended model
could also be used to study condensate formation in systems far from 
equilibrium. Another application might consider vortex dynamics at finite 
temperatures. A vortex in this case would be an `obstacle' in the mean-field
experienced by non-condensed particles, leading to scattering and hence to
a net force on the vortex. This should result in the expected drift of the 
vortex to the condensate edge, and allow direct determinations of vortex
lifetimes. Similar models could also facilitate a fully consistent description
of dissipative processes during probing of the condensate by a moving object 
\cite{jackson00}.

We would like to acknowledge valuable discussions with Jim McCann, and 
financial support from EPSRC.


\begin{references}
\bibitem{softer} A `softer' drive is also imposed by varying the trap 
 frequencies over several oscillations. We find
 that our results are largely independent of the form of the excitation, apart
 from a slight dependence on the oscillation amplitude for large values.
\end{references}
\end{document}